\documentclass{article}
\usepackage{arxiv}
\usepackage{hyperref}
\usepackage{latexsym}
\usepackage{url}
\usepackage[utf8]{inputenc}
\usepackage{booktabs}
\usepackage{amsmath}
\usepackage{amsfonts}
\usepackage{amssymb}
\usepackage[inline]{enumitem}
\usepackage{enumitem}
\usepackage{mathbbol}
\usepackage[T1]{fontenc}
\usepackage{mathtools}
\usepackage{multirow}
\usepackage{natbib}
\usepackage{rotating}
\usepackage{subcaption}
\usepackage{graphicx}
\usepackage{etaremune}
\usepackage{authblk}
\usepackage{tikz}

\newcommand{\eg}{\emph{e.g.}}

\title{Overview of the TREC 2025 Tip-of-the-Tongue track}

\author[1]{Jaime Arguello}
\author[2]{Fernando Diaz}
\author[3]{Maik Fröebe}
\author[2]{To Eun Kim}
\author[4]{Bhaskar Mitra}

\affil[1]{University of North Carolina, USA\\\texttt{\small jarguell@email.unc.edu}}
\affil[2]{Carnegie Mellon University, USA\\\texttt{\small diazf@acm.org, toeunk@cs.cmu.edu}}
\affil[3]{Friedrich-Schiller-Universität Jena, Germany\\\texttt{\small maik.froebe@uni-jena.de}}
\affil[4]{Independent Researcher, Canada\\\texttt{\small 
bhaskar.mitra@acm.org}}

\begin{document}
\maketitle

\begin{abstract}
Tip-of-the-tongue (ToT) known-item retrieval involves re-finding an item for which the searcher does not reliably recall an identifier. ToT information requests (or queries) are verbose and tend to include several complex phenomena, making them especially difficult for existing information retrieval systems. The TREC 2025 ToT track focused on a single ad-hoc retrieval task. 
This year, we extended the track to general domain and incorporated different sets of test queries from diverse sources, namely from the MS-ToT dataset, manual topic development, and LLM-based synthetic query generation.
This year, 9 groups (including the track coordinators) submitted 32 runs.
\end{abstract}
\begin{center}
    \textbf{Track website:} \url{https://trec-tot.github.io}
\end{center}
\section{Introduction}
\label{sec:intro}

Tip-of-the-tongue (ToT) known-item retrieval involves retrieving an item for which the searcher is unable to reliably recall an identifier---\eg, resolving the name of a movie for which the searcher does not recall the title or the name of a cast member.
ToT information requests (or queries) are verbose and include several complex phenomena.
First, they include information about the item itself (i.e., semantic memories) as well as the context in which the searcher last engaged with the item (i.e., episodic memories).
Additionally, they include language phenomena that simple keyword-matching algorithms are not equipped to handle, such as mentions of (un)certainty, exclusion criteria, relative comparisons, and false memories.
Such phenomena are not prevalent in verbose queries in other retrieval scenarios.

Current IR systems are not well-suited to resolve ToT information needs.
As evidence, a wide range of community Q\&A sites have emerged to help people resolve their ToT information needs through the help of other people.
Such Q\&A sites focus on domains such as movies\footnote{\url{https://www.reddit.com/r/tipofmytongue}}, books\footnote{\url{ https://www.goodreads.com/group/show/185-what-s-the-name-of-that-book}}, stories\footnote{\url{ https://scifi.stackexchange.com/questions/tagged/story-identification}}, and songs\footnote{\url{https://www.watzatsong.com/en}}. 

In the previous two years, the ToT track focused on a single ad-hoc retrieval task in a few selected domains (movie domain in 2023 ~\citep{arguello2023overview} and movie, landmark, and celebrity in 2024~\citep{arguello2024overview}).
This year, we extended the track to go beyond focusing on specific domains, with our test set covering $53$ different types of entities in total.
The TREC 2025 ToT track received 32 runs from 9 participant groups.

\section{Task description}
\label{sec:task}

The TREC 2025 ToT Track had a single ad-hoc retrieval task focusing on identification of objects or concepts in general domain.  
Participants were provided with a training set of 143 ToT queries from the Microsoft ToT Known-Item Retrieval Dataset for Movie Identification (MS-ToT) dataset~\citep{arguello:ToT}\footnote{\url{https://github.com/microsoft/Tip-of-the-Tongue-Known-Item-Retrieval-Dataset-for-Movie-Identification}}, which are sourced from the ``I Remember This Movie\ldots'' community Q\&A site.\footnote{\url{https://irememberthismovie.com}}
Participants were also provided three development sets.
The first two development sets originated from the same MS-ToT dataset, and served as the development set and the test set in the 2023 edition of the track.
The last development set originated from the TREC 2024 ToT track's test set, which includes synthetic queries in three domains: movie, landmark, celebrity.

Participants were provided with a test set of $622$ queries.
Of these, $172$ were new queries sampled from the MS-ToT Dataset (movie domain), and $150$ were human-elicited queries spanning three domains: movies, celebrities, and landmarks,  developed with the help of \emph{NIST} assessors.
The remaining $300$ test queries were LLM-generated (\emph{synthetic}) in the general domain using two different LLMs ($150$ by Llama-3.1-8B-Instruct and $150$ by GPT-4o).

The document corpus was composed of the subset of Wikipedia dumped in 2023 ($6,407,814$ articles).
The corpus contained the correct answer for every training, development, and test queries distributed to participants.
For each query, participants were asked to produce a ranking of (at most) 1000 document IDs from the Wikipedia corpus, with the correct answer ranked as high as possible.
Participants were allowed to use external resources such as Wikidata.
Runs were evaluated using several ranking metrics, with NDCG@1000 as the official metric.

As previously noted, ToT queries are verbose.
However, they contain complex phenomena that are not typically found in other retrieval scenarios involving verbose queries.
Such phenomena include: (1) memories about the item itself, (2) contextual memories, (3) false memories, (4) mentions of uncertainty, (5) mentions of previous failed attempts to re-find the item, (6) relative comparisons that require multi-hop reasoning to be useful, and (7) social niceties.

The following are examples of the four types of queries included in the test set: MS-ToT, NIST (human-elicited), Llama-generated, and GPT-generated.

\vfill
\noindent\fbox{
\parbox{\textwidth}{
\textbf{MS-ToT Movies}\\\\
I saw a clip on facebook from a black and white unknown charlie chaplin sort of movie in which there is sort of hilarious mayham when two people wearing a horse or mule costume are running around and people are trying to catch them and they kick a few individuals probably guards and finally run away in same horse costume riding a bicycle. Can anybody please tell me which movie was that which has got this scene. Thanks and regards\\\\
\textbf{Correct Answer: Another Fine Mess}\\
\textbf{Domain: movie}
}    
}

\vfill
\noindent\fbox{
\parbox{\textwidth}{
\textbf{NIST (human-elicited)}\\\\
so this guy is an actor. most recently though i've been seeing him on tv commercials for casinos maybe. and he's the guy, or one of two men, on a talk show with several women. it's not the talk show with whoopi goldberg i'm pretty sure. the tv commericals have something to do with games and money\\\\
\textbf{Correct Answer: Jerry O'Connell}\\
\textbf{Domain: celebrity}
}
}

\vfill
\noindent\fbox{
\parbox{\textwidth}{
\textbf{Synthetic (Llama-generated)}\\\\
I'm trying to remember this particular type of cloud that used to fascinate me when I spent summers on my grandparents' rural farm. They were these wispy things that showed up high up in the sky, sometimes days before a big storm rolled in. I remember their ethereal beauty, like delicate brushstrokes across the sky. They looked like wrinkles in the heavens. These clouds seemed to thicken and spread out as time passed, as if grains of diaphanous threads were being pulled tight. I recall teasing my siblings, daring them to touch the clouds, even though they were far out of reach. I think these clouds were connected to more massive cloud formations that came later. They predicted weather changes, but we never actually got rain from them themselves, weirdly enough. My grandfather said something about them being an early warning system for larger storms. His words are stuck in my head: ""A sign of the big ones coming, but not the ones to bring the water."" I don't know why these memories are coming up now, but I suppose someone here must know what I'm thinking of...\\\\
\textbf{Correct Answer: Cirrostratus fibratus}\\
\textbf{Domain: cloud type}
}    
}

\vfill
\noindent\fbox{
\parbox{\textwidth}{
\textbf{Synthetic (GPT-4o-generated)}\\\\
I remember this certain type of cheese from way back when I visited a region that bordered Switzerland, famous for its rolling hills and pastoral beauty. This cheese wasn't the superstar on the international stage, but it had a devoted following locally. It had this peculiar, almost melt-in-your-mouth quality once it was paired with something else in a rustic pot. The locals seemed to have a ritual around simmering it with a bit of liquidsometimes water, sometimes something creamierand they'd often toss in some spices or that pungent bulb that gives dishes a real bite. The aroma of it cooking was warm and inviting, kind of like a hug from a long-lost friend. I distinctly remember sitting with an old friend in a stone cottage, where the crackling fire enhanced the experience of this silky cheese concoction. The locals took pride in the tradition of how it was made. I guess what struck me was not just the unique taste and texture but also how the whole process seemed like a homage to their agricultural roots. Does anyone else recall a cheese with a story or ritual like this tethered to it? Maybe I'm just romanticizing it, but I'd love to identify this elusive fromage again. Any clues would be much appreciated! \\\\
\textbf{Correct Answer: Metton}\\
\textbf{Domain: cheese}
}    
}

\section{Data}
\label{sec:data}

\subsection{Corpus Construction}
\label{sec:data-corpus}

Participants were provided with a corpus of 6,407,814 Wikipedia articles associated (directly or indirectly) with Wikipedia categories relevant to our $53$ general domains.  The corpus contained the correct answer for all training, development, and test queries.  Each document in the corpus included the following fields:
\begin{itemize}[noitemsep,topsep=0pt]
\item \textbf{doc\_id:} Unique document identifier (Wikipedia page ID).
\item \textbf{url:} Wikipedia page URL.
\item \textbf{title:} Wikipedia page title.
\item \textbf{text:} Full text of the Wikipedia page. 
\end{itemize}

Section information was provided in case participants wanted to focus on only certain sections of the Wikipedia page.

\subsection{Human-Elicited ToT Query Collection}

The test set included 150 ToT queries elicited from NIST assessors.  These included ToT queries in the domains of Celebrities, Landmarks, and Movies.  These queries were elicited from NIST assessors using images.  For Celebrities and Landmarks, we sampled images directly from Wikipedia.  For Movies, many Wikipedia images included the movie title.  Therefore, we sampled movie images from IMDB and TMDB (linked to from a Wikipedia page).

Some manual filtering was necessary.  For the Celebrities domain, we filtered images with multiple individuals.  For Movies, we filtered images with nudity and excessive gore.  For Landmarks, we filtered images with the landmark name.  Ultimately, we sampled 1,946 images for Celebrities, 330 images for Landmarks, and 1,687 images for Movies.

\begin{figure*}[]
\centering
\includegraphics[trim=0 160 0 120, clip, width=\textwidth]{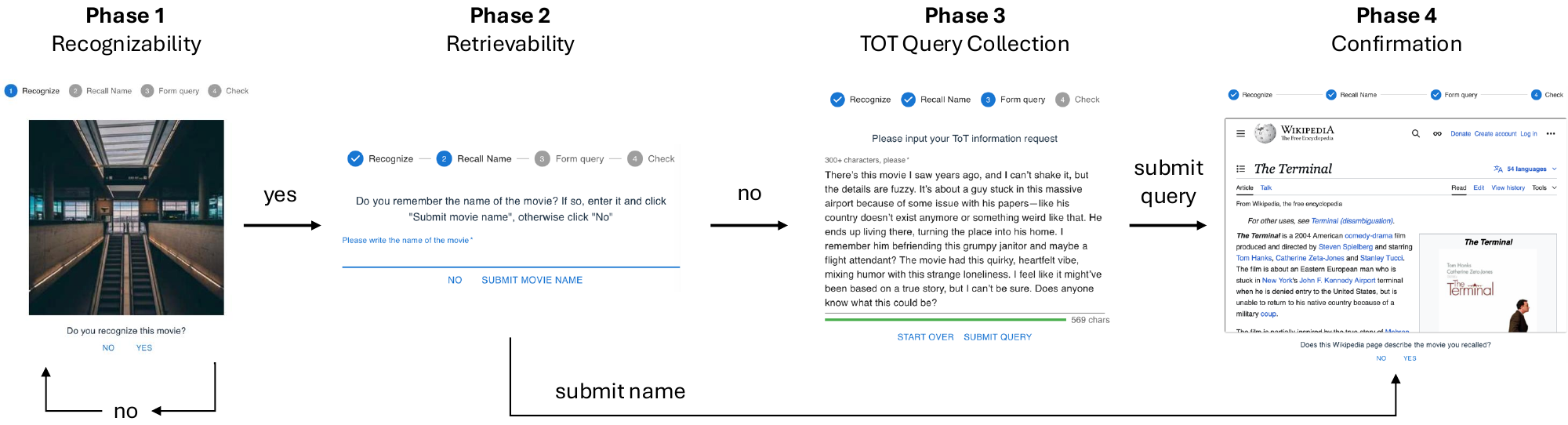}
\caption{Flowchart of the Human ToT Query Elicitation Interface.}\label{fig:interface-flowchart}
\end{figure*}

The query elicitation process proceeded in four phases and is depicted in Figure~\ref{fig:interface-flowchart}.  During Phase 1 (Recognizability), the NIST assessor was displayed an image.  Images were selected from our three domains in a round-robin fashion.  The system asked the NIST assessor whether they recognized the entity in the image.  If the NIST assessor did not recognize the entity, the system returned to Phase 1 with a new image.  If the NIST assessor recognized the entity, the system proceeded to Phase 2. 

During Phase 2 (Retrievability), the system asked the NIST assessor if they recalled the name of the entity (i.e., Celebrity, Landmark, or Movie) depicted in the image.  If the NIST assessor indicated ``yes'', the system proceeded to Phase 4, which displayed the Wikipedia page associated with the image and asked the NIST assessor to confirm whether the recalled entity was correct.  If the NIST assessor indicated ``no'', they proceeded to Phase 3.

During Phase 3 (ToT Query Elicitation), the system prompted the NIST assessor to produce a ToT query.  To encourage NIST assessors to submit verbose ToT queries, a progress bar started as red when the query was less than 200 characters long, turned yellow on exceeding 200 characters, and turned green once exceeding 300 characters.

Finally, during Phase 4 (Entity Confirmation), the system displayed the Wikipedia page of the entity and asked the assessor to confirm whether it corresponded to the entity they recognized (Phase 1) but could not recall (Phase 2).

For the human-elicited ToT query to be \emph{valid}, the NIST assessor had to: (1) recognize the image (Phase 1); (2) be unable to recall the name from memory (Phase 2), and (3) confirm that the Wikipedia page associated with the entity matched the entity they recognized but could not recall (Phase 4).  Ultimately, we collected 53 valid queries for Celebrities, 47 for Landmarks, and 50 for Movies.

More detailed information on the collection process and data statistics can be found from \cite{he2025totSimulatedEval}.

\subsection{Synthetic Query Generation}
\label{sec:data-query}

In addition to queries from the MS-ToT Dataset~\citep{arguello:ToT} and Human-elicited collection \citep{he2025totSimulatedEval}, we generated synthetic queries using two large language models.

To generate synthetic ToT queries, we first sampled random Wikipedia articles from our corpus (Section \ref{sec:data-corpus}). Each sampled article’s title and summarized body text were then used to seed a large language model (LLM) through a domain-agnostic prompt (Figure \ref{fig:prompt}) adapted from the method introduced by \cite{he2025totSimulatedEval}.
Using the infobox metadata, we randomly selected 50 domains from the corpus to serve as test categories. For each domain, we sampled 6 Wikipedia pages and summarized their content using GPT-4o. Each summarized page was then used to construct a synthetic query generation prompt. Three of these prompts were passed to Llama-3.1-8B-Instruct and the remaining three to GPT-4o, yielding three ToT entities per LLM in each of the 50 domains. This process resulted in 150 ToT entities per LLM, producing a total of 300 synthetic ToT queries.

\begin{figure}
\noindent\fbox{
\parbox{\textwidth}{
\textbf{System Prompt:}\\
You are a user on an online forum and want to ask what is on the tip of your tongue.\\
\textbf{Main Prompt:}\\
Let's do a role play. You are someone who knew about \{ToTObject\} from a long time ago and are now struggling to recall its exact name. 
You're trying to remember it by posting vague description of it on an online forum like Reddit. 
Generate a post of about 200 words that describes your partial and hazy memory of \{ToTObject\}. 
Your post should avoid mentioning the exact name and make it genuinely hard for others (and search engines) to identify the correct answer. 
I will provide you with some basic information about \{ToTObject\}, which you should use to guide your memory, following the rules below.\\

Information about \{ToTObject\}: \{Summarized Wikipedia Page\}\\

Guidelines:\\
MUST FOLLOW:\\
1. Reflect the imperfect nature of memory using expressions of uncertainty or mixed-up details, without directly stating "I'm not sure if this is true".\\
2. Do not include the name of the entity or any exact identifiers (e.g., titles, brand names, person names).\\
3. Describe components, features, or roles in indirect or descriptive ways.\\
4. Use a conversational tone that feels natural, avoiding formal writing structures.\\
5. Include vivid but ambiguous sensory or contextual details that spark curiosity and reflect partial recall.\\
6. Do not copy any example phrases verbatim; create a unique and realistic expression of memory.\\
7. Skip formal greetings like "Hello" or "Hey everyone" --- begin directly with your memory.\\

COULD FOLLOW:\\
1. Set the scene with a personal anecdote about when or where you encountered the entity, avoiding clichés like "When I was young."\\
2. Focus on how the entity made you feel or what stood out to you emotionally or physically.\\
3. Make subtle comparisons to similar experiences or other entities without explicitly naming them.\\
4. Include one or two plausible but incorrect details to reflect natural memory distortions.\\
5. Mention specific parts, events, or moments associated with the entity, described indirectly.\\
6. Use unique or less obvious time/place references (instead of "10 years ago" or "on TV").\\
7. End with an open-ended question or comment that invites others to help you figure it out.\\

Generate a post based on these guidelines.
}    
}
\caption{Prompt to Synthesize ToT Queries in General Domain.}
\label{fig:prompt}
\end{figure}


\subsection{External Sources}

Track participants were permitted and encouraged to leverage external sources beyond the Wikipedia corpus provided.  However, because 172 of our test queries originated from the MS-ToT Dataset, participant groups were cautioned to not tune/train their systems using data from the MS-ToT Dataset or data scrapped from the ``I Remember This Movie\ldots'' community Q\&A site.


\section{Results and Analysis}
\label{sec:result}

\subsection{Participation}
\label{sec:results-participation}
The TREC 2025 ToT track received a total of $32$ submissions from $9$ groups, including three baseline submissions from track coordinators.
Same as last year, participants were asked to report if they were certain that test data was not used to train their models.
Excluding the baseline runs, this year two participating groups answered in the affirmative for all their submitted runs which corresponded to $11$ runs in total including two baseline runs.
While this is more than last year, it still continues to reflect how it is becoming increasingly difficult to attest to such a claim in the absence of transparency around the data used for pretraining different LLMs that participants employed in producing their runs.

\noindent \textbf{External Data Usage} The submission form had an additional field for reporting if external data were also used.
$18$ runs, including all the baselines, used only the provided training data.
Six runs used external datasets in addition to the provided training datasets, and eight runs did not use the provided training data and leveraged other datasets exclusively.
The ``Training data'' column in Table \ref{tab:results} reports what data was leveraged for individual runs.

\noindent\textbf{Baseline usage}
The track coordinators submitted two BM25 baselines (with Anserini~\cite{yang:2017} and PyTerrier~\cite{macdonald:2021}) and a dense-retrieval baseline (with Lightning~IR~\cite{schlatt:2025a}), which were made available to participants.\footnote{Please see \url{https://github.com/TREC-ToT/bench/} for more details about the baselines}
Baselines are marked with an asterisk in Table \ref{tab:results}.
Of the $29$ submissions (excluding the baseline runs themselves), seven submissions reported to have used the baseline runs either as re-ranking candidates or as negative samples.
Runs which leveraged the baseline runs are marked in the column ``Used baseline'' in Table \ref{tab:results}.
None of the top-four runs this year used the baseline runs provided.

\subsection{Overall results}
\label{sec:results-overall}

\begin{table}[t]
    \caption{Summary of results. Best scores are in bold. The baselines are marked with an asterisk.}
    \label{tab:results}
    \begin{center}
    \resizebox{\textwidth}{!}{
  \begin{tabular}{lllccccc}
			run\_id & group\_id & data\_category & Used baseline & ndcg\_cut\_10 & recip\_rank & recall\_1000 & ndcg\_cut\_1000 \\
			\hline
			scrb-tot-04 & SRCB & Official and other &  & \textbf{0.6576} & \textbf{0.6258} & \textbf{0.9051} & \textbf{0.6824} \\
			scrb-tot-03 & SRCB & Official and other &  & 0.6547 & 0.6211 & \textbf{0.9051} & 0.6787 \\
			scrb-tot-02 & SRCB & Official and other &  & 0.6449 & 0.6097 & \textbf{0.9051} & 0.6700 \\
			scrb-tot-01 & SRCB & Official and other &  & 0.6162 & 0.5844 & 0.8955 & 0.6458 \\
			gmn-rerank-500 & DS@GT & Official only & \checkmark & 0.3831 & 0.3674 & 0.6559 & 0.4106 \\
			gm27q-comb-500 & DS@GT & Official only & \checkmark & 0.3619 & 0.3436 & 0.5884 & 0.3848 \\
			webis-bm25-gpt-oss & webis & Other only & \checkmark & 0.3078 & 0.2843 & 0.7637 & 0.3689 \\
			gm27q-LMART-1000 & DS@GT & Official only & \checkmark & 0.3046 & 0.2867 & 0.6109 & 0.3339 \\
			gemini-retrieval & DS@GT & Official only &  & 0.2949 & 0.2790 & 0.3505 & 0.2962 \\
			runid4 & ufmg & Official only &  & 0.2136 & 0.1986 & 0.6752 & 0.2771 \\
			webis-bm25-llama & webis & Other only & \checkmark & 0.1968 & 0.1790 & 0.6785 & 0.2619 \\
			runid1 & ufmg & Official only &  & 0.1771 & 0.1625 & 0.6752 & 0.2462 \\
			runid3 & ufmg & Official only &  & 0.1588 & 0.1412 & 0.6495 & 0.2218 \\
			dgMxbaiL01 & dgthesis & Official and other & \checkmark & 0.1620 & 0.1505 & 0.4855 & 0.2071 \\
			bm25\_hedge\_aware & uva illc & Other only &  & 0.1257 & 0.1136 & 0.5048 & 0.1759 \\
			bm25\_negations & uva illc & Other only &  & 0.1223 & 0.1114 & 0.4855 & 0.1700 \\
			pyterrier-bm25* & coordinators & Official only &  & 0.1223 & 0.1114 & 0.4855 & 0.1700 \\
			runid2 & ufmg & Official only &  & 0.1043 & 0.0964 & 0.5531 & 0.1640 \\
			top\_model\_dense & DS@GT & Official only &  & 0.1088 & 0.0960 & 0.5096 & 0.1620 \\
			bm25-porterblk-test & DUTH\_XANTHI & Official only &  & 0.1021 & 0.0946 & 0.4807 & 0.1537 \\
			lex-stronger-testv2 & DUTH\_XANTHI & Official only &  & 0.0992 & 0.0899 & 0.5064 & 0.1537 \\
			bge-m3 & DS@GT & Official only &  & 0.0852 & 0.0743 & 0.5498 & 0.1492 \\
			rm3\_hedges & uva illc & Other only &  & 0.0954 & 0.0821 & 0.5016 & 0.1481 \\
			rm3\_hedge\_neg & uva illc & Other only &  & 0.0954 & 0.0821 & 0.5016 & 0.1481 \\
			lambdamart-rerank & DS@GT & Official and other & \checkmark & 0.0769 & 0.0601 & 0.6109 & 0.1452 \\
			lex-stronger-test & DUTH\_XANTHI & Official only &  & 0.0939 & 0.0844 & 0.4502 & 0.1398 \\
			rm3\_negations & uva illc & Other only &  & 0.0846 & 0.0725 & 0.4807 & 0.1371 \\
			anserini-bm25* & coordinators & Official only &  & 0.0763 & 0.0724 & 0.2749 & 0.1032 \\
			llama\_norm\_fusion\_v2 & mst & Official only &  & 0.0466 & 0.0383 & 0.2958 & 0.0829 \\
			lightning-ir-dense* & coordinators & Official only &  & 0.0189 & 0.0160 & 0.2653 & 0.0514 \\
			llama\_norm\_fusion\_z & mst & Official only &  & 0.0005 & 0.0002 & 0.0032 & 0.0008 \\
			bm25\_hedges\_neg & uva illc & Other only &  & 0.0000 & 0.0000 & 0.0000 & 0.0000 \\
		\end{tabular}
    }	
	\end{center}
\end{table}

\begin{figure}[t]
    \centering
    \includegraphics[width=\textwidth]{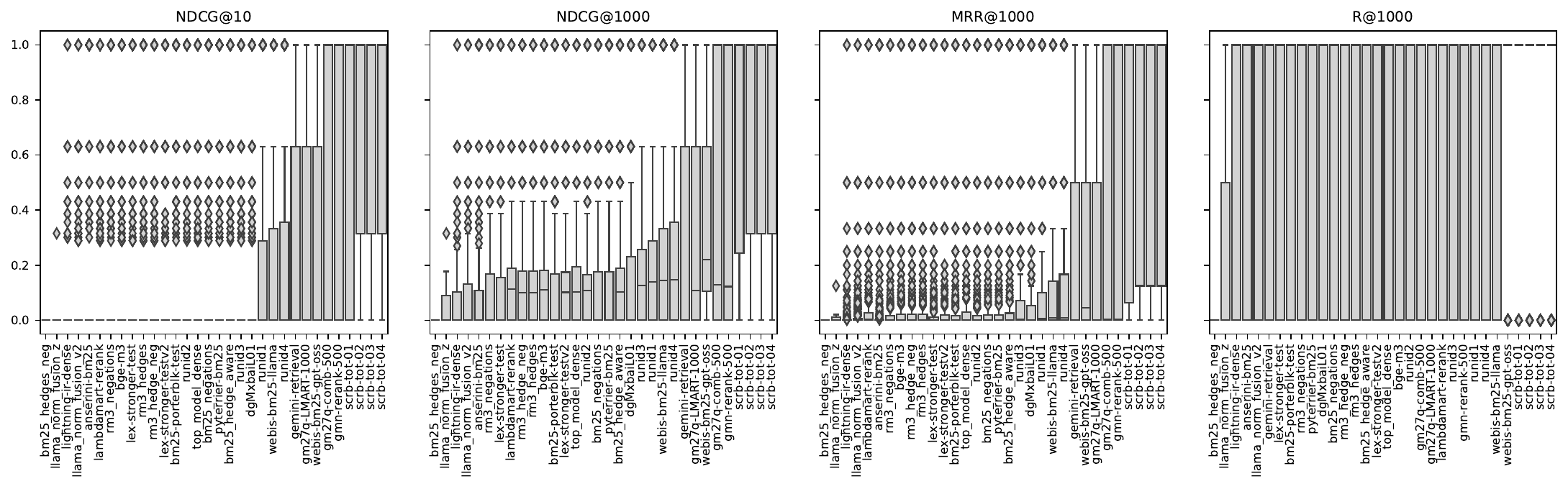}
    \caption{Metric distribution by run (sorted by mean) as boxplot showing the 25th, 50th, and 75th~quartiles.}
    \label{fig:metric-distribution-by-run}
\end{figure}

Table~\ref{tab:results} contains the results of all submitted runs including the baselines.
This year we continued to observe a wide spread of NDCG@1000 scores, as well as other metric scores, across the different runs.
We plotted the distribution of metrics across different runs in Figure~\ref{fig:metric-distribution-by-run}.
The runs along the x-axis were sorted by their median score.
Of the two submitted baselines, the PyTerrier-based BM25 baseline (\textit{pyterrier-bm25}) achieved the best performance.

\begin{figure}[t]
    \centering
    \includegraphics[width=0.8\linewidth]{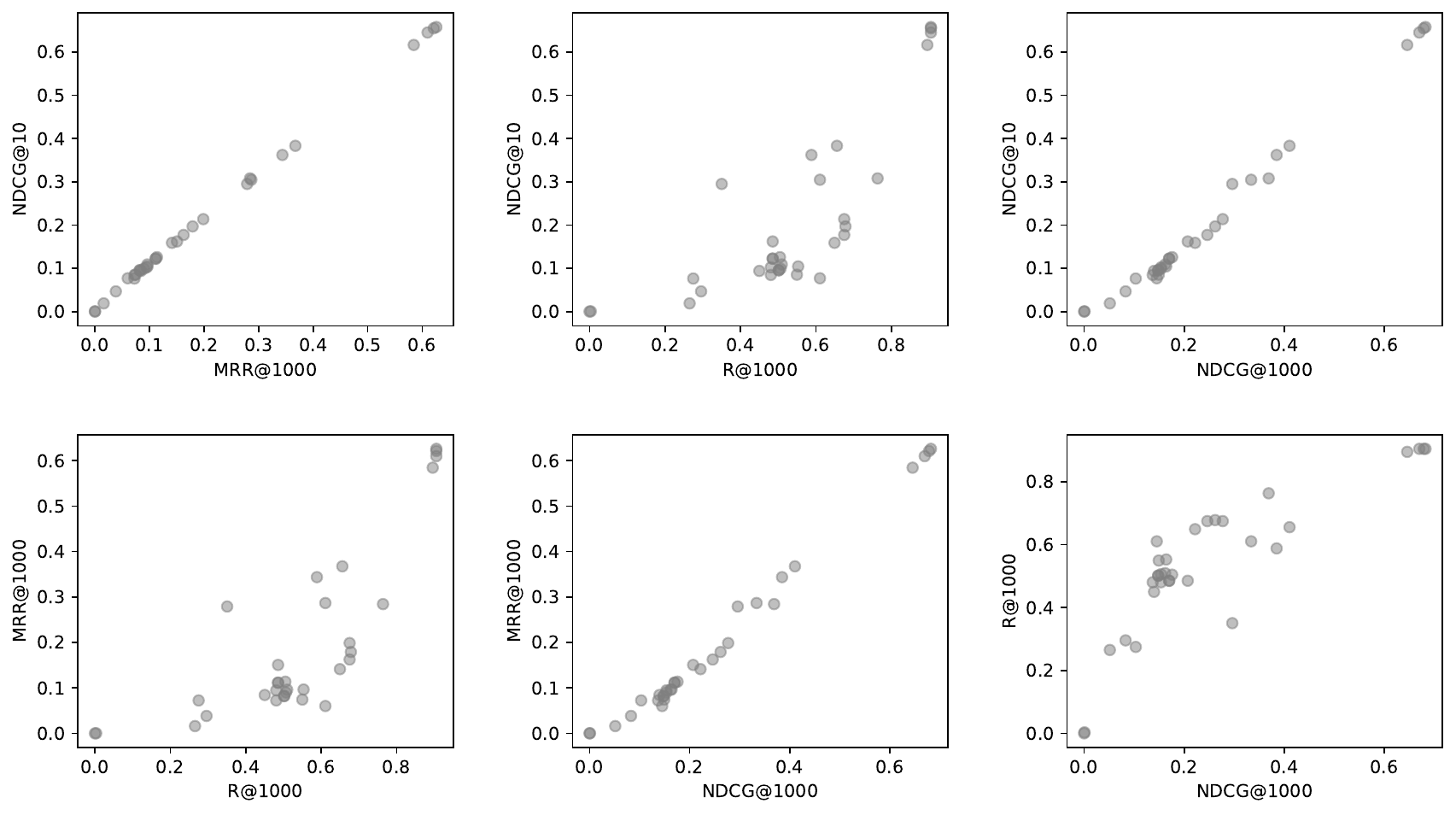}
    \caption{Metric correlations. Each dot represents the mean performance of a run, with the different metrics on the axes. We observe strong correlations between NDCG@10, NDCG@1000, and MRR. Recall also seems to correlate but less strongly with the other three metrics.}
    \label{fig:metric-correlation}
\end{figure}

\begin{figure}[t]
    \centering
    \includegraphics[width=\textwidth]{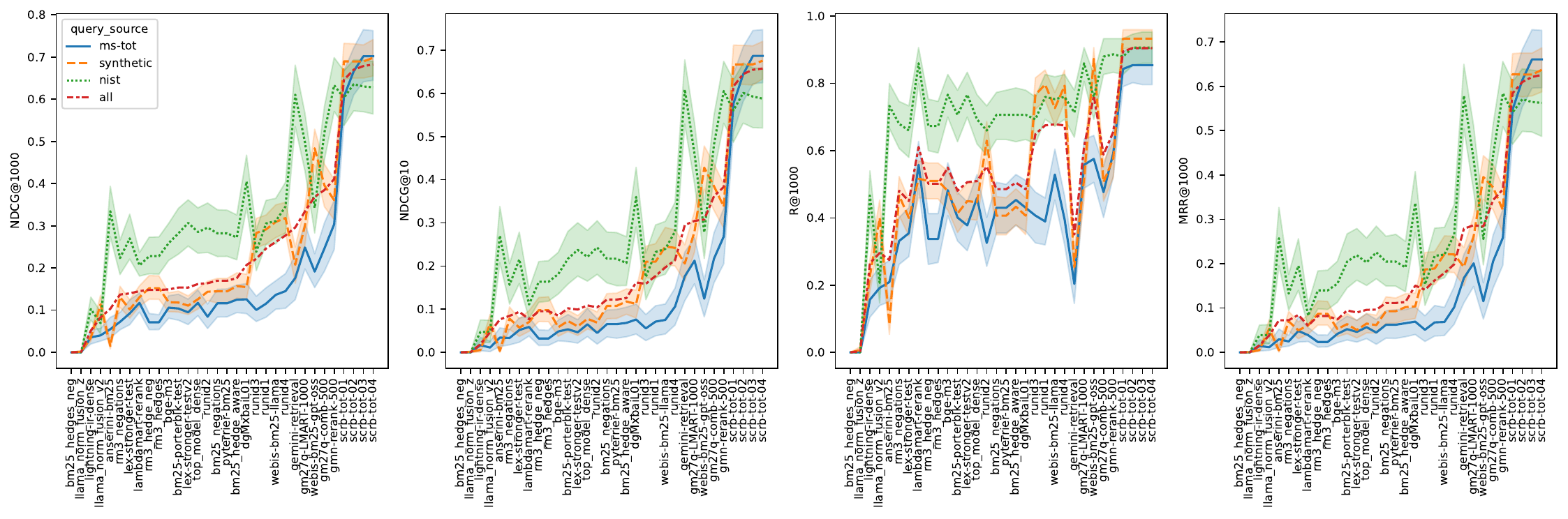}
    \caption{The mean effectiveness for all runs as nDCG@1000 (our sorting criteria), nDCG@10, Recall@1000, and MRR@1000. We show the mean for the three query types (ms-tot, synthetic, and NIST queries) and on all queries.}
    \label{fig:metric-distribution-by-team-and-query-source-mean}
\end{figure}

\begin{figure}[t]
    \centering
    \includegraphics[width=\textwidth]{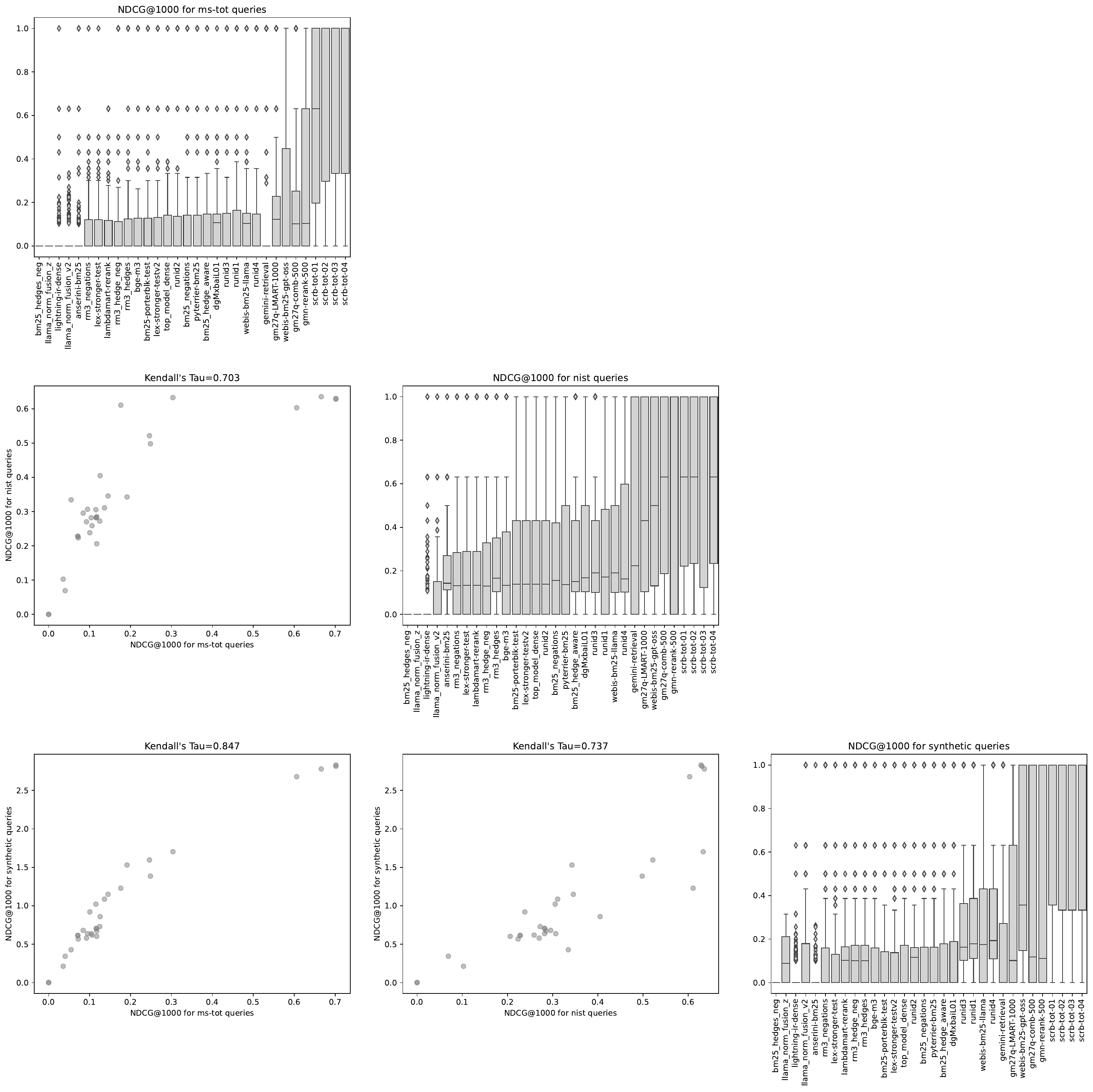}
    \caption{System performance correlation across query types.}
    \label{fig:metric-correlation-sources}
\end{figure}

The different performance metrics for the submitted runs seem to be well correlated to each other, except R@1000 which seemed to be less-correlated with other metrics this year, as shown in Figure~\ref{fig:metric-correlation}.
This year, the performance of participating systems on synthetic queries and NIST queries were more comparable to their performance on the MS-ToT movie queries as shown in Figure~\ref{fig:metric-distribution-by-team-and-query-source-mean}.
The system performances on MS-ToT queries and synthetic queries are fairly well correlated (Kendall's Tau=0.847), while the performance on NIST queries was relatively less correlated with that on MS-ToT queries (Kendall's Tau=0.703) and synthetic queries (Kendall's Tau=0.737), as shown in Figure~\ref{fig:metric-correlation-sources}.
This finding supports the use of diverse query generation approaches, including synthetic query generation and manual topic development, for retrieval evaluation for ToT information needs.

\begin{figure}[ht]
    \centering
    \includegraphics[width=\linewidth]{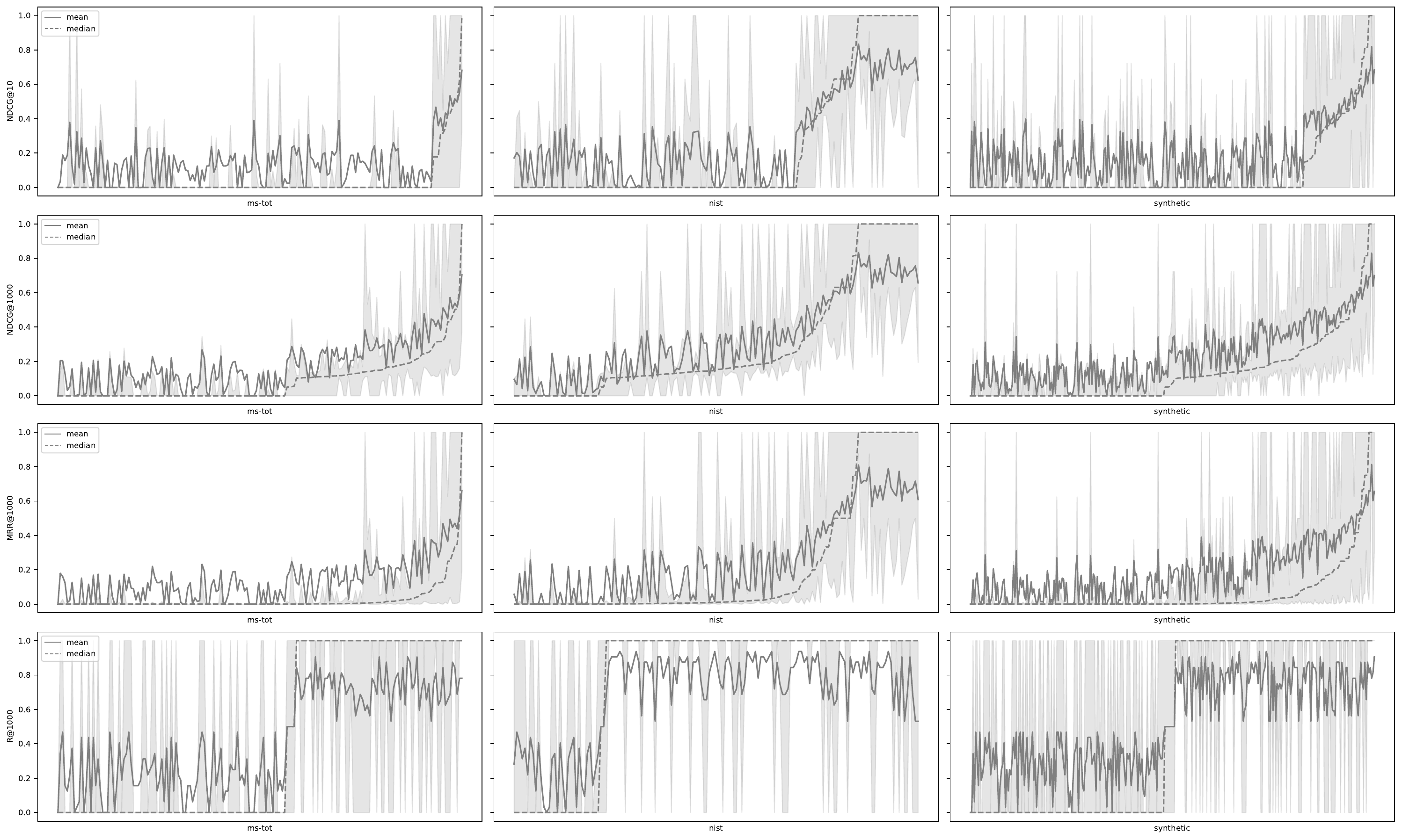}
    \caption{Metric distribution by query.}
    \label{fig:metric-distribution-by-query}
\end{figure}

We plotted the performance of different runs on different types of test queries in Figure \ref{fig:metric-distribution-by-query}, where the x-axis is sorted by median score.
From this plot, we can see that some queries were easier to resolve compared to others, and on average the MS-ToT movie queries were more difficult than NIST and synthetic queries.
NIST queries in particular seem to include a larger proportion of ``easy'' queries whose median system performances are close to the metric maximum.

\begin{figure}[ht]
    \centering
    \includegraphics[width=\textwidth]{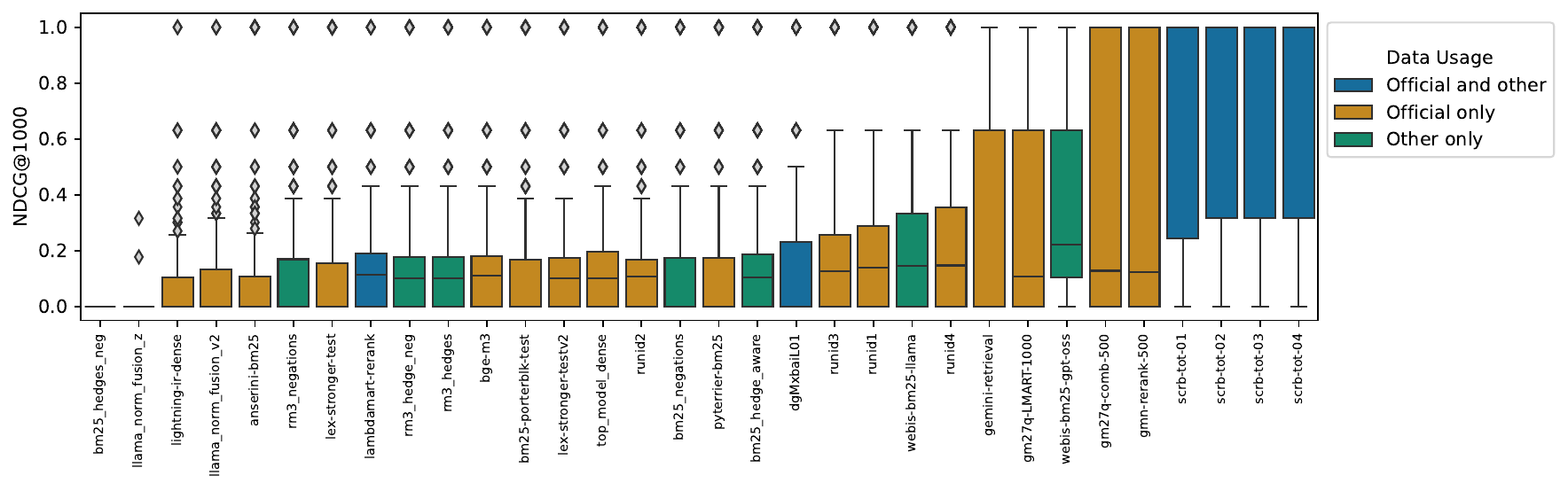}
    \caption{Boxplot showing the distribution of NDCG@1000 scores for different runs, with colors indicating if external data was used to produce the run (sorted by mean nDCG@1000).}
    \label{fig:data-usage-ndcg-runs}
\end{figure}

\noindent \textbf{Usage of external data} As mentioned before,  participants self-reported if external data was being used.
We plotted the runs colored by whether external data was used in Figure \ref{fig:data-usage-ndcg-runs}.
The top four runs leveraged a combination of provided and other datasets for training.
The next two runs used only the provided training data, followed by a run that exclusively uses other training data than the one provided.

\begin{figure}[t]
    \centering
    \includegraphics[width=\linewidth]{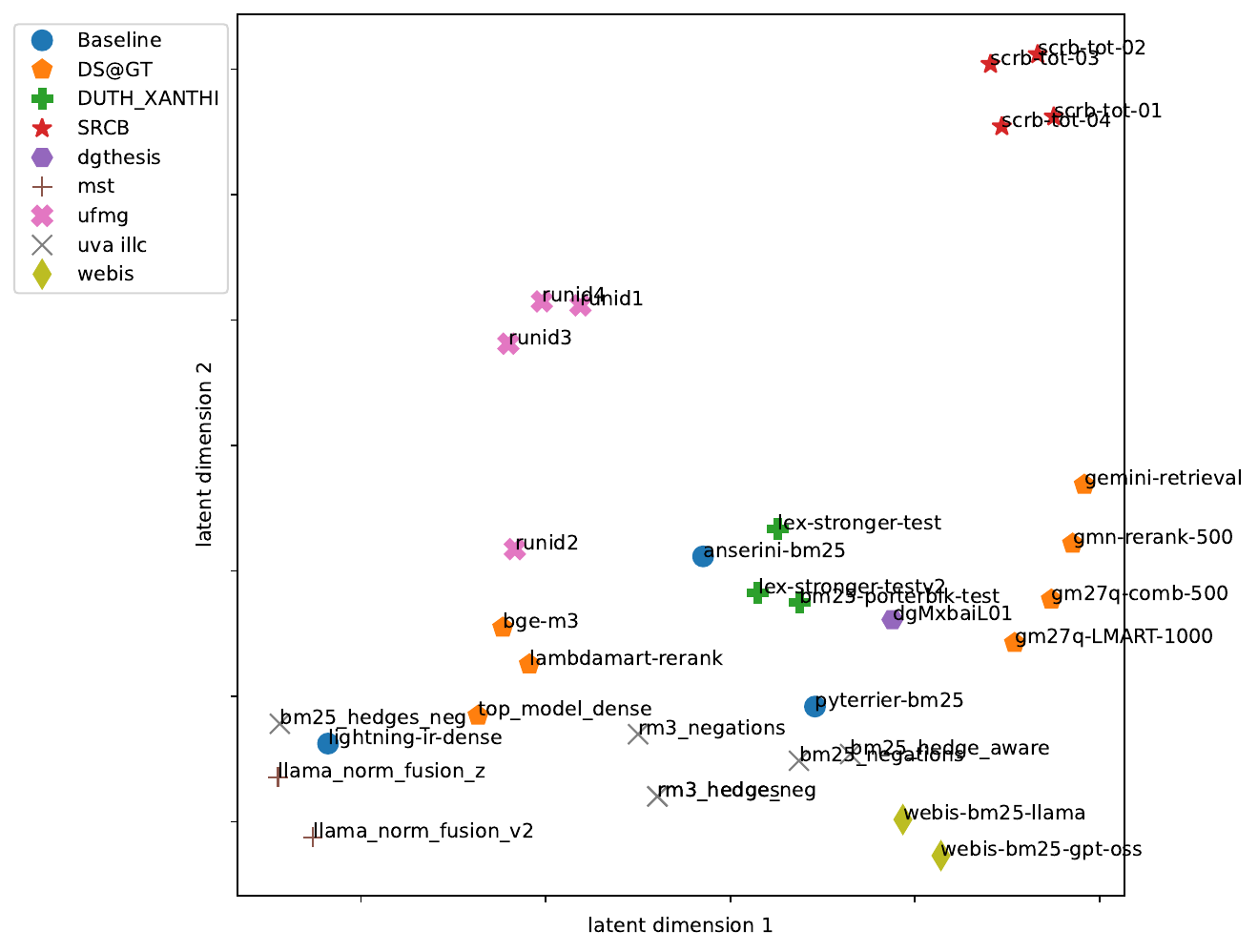}
    \caption{TSNE plot based on the NDCG@1000 scores of different runs.}
    \label{fig:tsne-groups}
\end{figure}

\noindent \textbf{TSNE} We plotted a TSNE plot of the runs in Figure \ref{fig:tsne-groups}.
The TSNE reduction was performed on the NDCG@1000 scores for each topic.
We can see that runs from the same group often cluster together.

\bibliographystyle{abbrvnat}
\bibliography{bibtex}

\end{document}